\documentclass{elsart}

\usepackage{amssymb, amsmath}
\usepackage{bbm}
\usepackage{subfigure}
\usepackage{here}
\usepackage{graphicx}
\usepackage[dvips]{color}

\newcommand{\Psihat}{\hat{\Psi}}
\newcommand{\Hhat}{\hat{H}}
\newcommand{\kron}[2]{\delta_{#1,#2}}
\newcommand{\Gamtens}[2]{\Gamma_{\alpha_{#1} \alpha_{#2}}^{[#2]n_{#2}}}
\newcommand{\lamtens}[1]{\lambda_{\alpha_{#1}}^{[#1]}}
\newcommand{\lamtilde}[1]{\tilde{\lambda}_{\tilde{\alpha}_{#1}}^{[#1]}}

\begin{document}

\begin{frontmatter}



\title{Ultracold Atoms in 1D Optical Lattices:\\  Mean Field, Quantum Field, \\ Computation, and Soliton Formation}


\author{R. V. Mishmash} and
\author{L. D. Carr}

\address{Department of Physics, Colorado School of Mines,
Golden, CO, 80401}

\begin{abstract}
In this work, we highlight the correspondence between two descriptions of a
system of ultracold bosons in a one-dimensional optical lattice potential: (1)
the discrete nonlinear Schr\"{o}dinger equation, a discrete mean-field theory,
and (2) the Bose-Hubbard Hamiltonian, a discrete quantum-field theory. The
former is recovered from the latter in the limit of a product of local coherent
states.  Using a truncated form of these mean-field states as initial
conditions, we build quantum analogs to the dark soliton solutions of the
discrete nonlinear Schr\"odinger equation and investigate their dynamical
properties in the Bose-Hubbard Hamiltonian.  We also discuss specifics of the
numerical methods employed for both our mean-field and quantum calculations,
where in the latter case we use the time-evolving block decimation algorithm
due to Vidal.
\end{abstract}

\end{frontmatter}

% main text
\section{Introduction}

Ultracold atoms in optical lattices have recently been at the center of
exciting research by both experimental and theoretical physicists alike.  In
these systems, experimentalists are given an unprecedented amount of control
over system parameters, and theorists are able to accurately describe the
physics with relatively simple, clean models.  However, a full quantum
many-body treatment of the problem is computationally challenging due to the
exponential growth of the Hilbert space with the system size. There do exist
numerous numerical methods, such as quantum Monte Carlo
\cite{Foulkes07_RevModPhys_73_33} and density matrix renormalization group
\cite{Schollwock05_RevModPhys_77_259}, that can accurately calculate the ground
state of the governing quantum Hamiltonian. On the other hand, for systems not
exhibiting strong correlations, one can safely employ an appropriate mean-field
theory and describe the many-body system semiclassically, resulting in a much
more tractable mathematical problem. Also, before the realization of ultracold
atoms, dynamical properties of the many-body lattice problem saw little
attention; however, far-from-equilibrium quantum dynamics has recently become a
hot topic of research \cite{Kinoshita06_Nature_440_04693,
Kollath07_PRL_95_180601}. Numerical work in this area has been spurred by the
advent of a newly available algorithm for simulating the quantum dynamics of
one-dimensional lattice Hamiltonians \cite{Vidal04_PRL_93_040502}.

In this paper, we study dynamical properties of ultracold bosons on
one-dimensional (1D) optical lattices.  Our presentation is organized as
follows. First, we elucidate the correspondence between the mean-field and
quantum many-body theories in the case of the tight-binding approximation on a
lattice. Second, we discuss our numerical methods, including an implementation
of the time-evolving block decimation algorithm to simulate the quantum
dynamics \cite{Vidal04_PRL_93_040502}. Finally, we simulate and analyze the
time evolution of dark solitons under both the semiclassical and quantum
equations of motion and highlight the observed differences between the two
theories.

\section{Quantum-Mean Field Correspondence}

\subsection{Mean-Field Theory}

The statics and dynamics of a weakly interacting Bose gas at zero temperature
in free space, or in the more experimentally relevant geometry of a harmonic
trap, are well-described by the Gross-Pitaevskii equation (GP), a.k.a., the
nonlinear Schrodinger equation (NLS) \cite{Dalfovo99_RevModPhys_71_463}.  For
these simple geometries, even in the quasi-1D regime, such a mean-field
approach is appropriate given that there is negligible depletion out of the
condensed mode. However, quantum fluctuations cannot be ignored throughout time
evolution for certain excited condensate states such as the dark soliton
\cite{Dziarmaga02_PRA_66_043615, Dziarmaga03_JPhysB_36_1217}. Mean-field theory
treatments have been generalized to lattice geometries
\cite{Bronski01_PRL_89_1402} in which case the lattice soliton solutions of the
NLS have been mapped out in detail \cite{Louis03_PRA_67_013602,
Efremidis03_PRA_67_063608}. However, such analysis assumes that the lattice
height is sufficiently low so that the gas still exhibits nearly ideal
Bose-condensation, i.e., negligible depletion out of the single boson
configuration as assumed by the NLS.  When using an unperturbed mean-field
theory to describe the system, one ideally uses the continuous NLS with an
external lattice potential; however, coupled-mode theory can be employed for a
shallow lattice \cite{Efremidis03_PRA_67_063608} or, in the other extreme, a
single-band tight-binding approximation can be used for a deep lattice
\cite{Trombettoni01_PRL_86_2353}.

In the latter case of the tight-binding approximation, assuming 1D from here
on, the full condensate wave function $\Phi(x,t)$ is expanded in a basis of
localized condensate wave functions $\phi(x-x_i)$ each centered at site $i$ in
the lowest Bloch band of the lattice. That is, $\Phi(x,t)=\sum_i \psi_i(t)
\phi(x-x_i)$. It is then assumed that the local condensate wave functions are
sufficiently localized in each well, in which case we obtain the well-known
discrete nonlinear Schr\"{o}dinger equation (DNLS):
\begin{equation}
i\hbar\partial_t\psi_k = -J\left (\psi_{k+1} + \psi_{k-1}\right) + U|\psi_k|^2
\psi_k + \epsilon_k\psi_k, \label{eqn:dnlse}
\end{equation}
where $\psi_k=\psi_k(t)$ is the dimensionless c-number that weights the
localized condensate wave function at the $k$th lattice site, and $\epsilon_k$
is a local external potential different from the lattice potential. The
coupling parameter $J$ and the nonlinearity $U$ can be written explicitly in
terms of overlap integrals involving the system parameters and the localized
condensate wave functions $\phi(x-x_i)$ \cite{Trombettoni01_PRL_86_2353}.

\subsection{Quantum Many-Body Theory}\label{sec:qmbt}

We now turn to the full quantum many-body description of the problem in the
regime of optical lattice depths where the above mean-field tight-binding
description is appropriate, and beyond. As was demonstrated by Jaksch \emph{et
al.} \cite{Jaksch98_PRL_81_3108}, a system of weakly interacting ultracold
bosons loaded in an optical lattice potential is an almost perfect realization
of the Bose-Hubbard Hamiltonian (BHH), a model introduced to the condensed
matter community almost ten years earlier by Fisher \emph{et al.}
\cite{Fisher89_PRB_40_546}.

To derive the BHH, we start with the 1D continuous many-body Hamiltonian in
second quantization for two-body interactions:
\begin{align} \Hhat = & \int dx \Psihat^\dagger(x)
\left[-\frac{\hbar^2}{2m}\frac{\partial^2}{\partial x^2} + V_\mathrm{ext}(x) \right]\Psihat(x) \nonumber \\
& + \frac{1}{2} \int dx \int dx'
\Psihat^\dagger(x)\Psihat^\dagger(x')V_\mathrm{int}(x-x')
\Psihat(x')\Psihat(x)\,. \label{eqn:contHami}
\end{align}
We then expand the bosonic field operator $\Psihat(x)$, which destroys a
particle at position $x$, in a lowest Bloch band Wannier basis as
$\Psihat(x)=\sum_i \hat{b}_i w(x-x_i)$, where the operator $\hat{b}_i$ is
defined to destroy a particle in the localized Wannier wave function
$w(x-x_i)$.  This step is analogous to expansion of the full condensate wave
function in a localized basis when discretizing the continuous NLS to obtain
the DNLS. Then, as is done in the derivation of the continuous NLS, we assume
the two-body interaction potential to be of the contact form, i.e.,
$V_\mathrm{int}(x-x_i)= g\delta(x-x_i)$, where $g$ is proportional to the
$s$-wave scattering length of the atoms.  We also invoke the tight-binding
approximation by assuming that the lattice is deep enough to obtain
sufficiently localized Wannier functions. This allows us to discard all terms
except those involving nearest-neighbor hopping and on-site interactions. After
making these assumptions, we arrive at the familiar BHH:
\begin{equation} \Hhat = -J
\sum_{i=1}^{M-1}(\hat{b}_{i+1}^\dagger \hat{b}_{i} + \mathrm{h.c.}) +
\frac{U}{2} \sum_{i=1}^{M} \hat{n}_i (\hat{n}_i-\hat{\mathbbm{1}}) +
\sum_{i=1}^M \epsilon_i \hat{n}_i, \label{eqn:bhh}
\end{equation}
where $\hat{b}_i$ and $\hat{b}^\dagger_i$ are destruction and creation
operators at site $i$ that obey the usual bosonic commutation relations, and
$\hat{n}_i\equiv \hat{b}^\dagger_i\hat{b}_i$ is the number operator which
counts the number of bosons at site $i$.  Equation (\ref{eqn:bhh}) assumes box
boundary conditions on a lattice containing $M$ sites. The coefficients $J$,
$U$, and $\epsilon_i$ can be calculated exactly in terms of the localized
single-particle wave functions and other parameters
\cite{Jaksch98_PRL_81_3108}.  $J$ is the nearest-neighbor hopping coefficient,
$U$ is the on-site interaction energy, and $\epsilon_i$ is an external
potential. The ratio $U/J$ is an important parameter that determines the
relative contribution from each term.  For a shallow lattice, the hopping term
dominates, whereas for a deep lattice the interaction term dominates.  However,
we note that $U$ is not completely dependent on the lattice geometry since the
$s$-wave scattering length can be varied independently via a Feshbach
resonance.
%\begin{align}
%J & = -\int dx w^\ast(x)\left[-\frac{\hbar^2}{2m}\frac{\partial^2}{\partial x^2} + V_\mathrm{lattice}(x)\right]w(x-a), \\
%U & = g\int dx |w(x)|^4, \\
%\epsilon_i & = \int dx V_\mathrm{ext}(x) |w(x-x_i)|^2\approx
%V_\mathrm{ext}(x_i),
%\end{align}
%where $a$ is the lattice constant and the external potential has
%been split into two terms, one for the lattice potential and one
%for an arbitrary external potential:
%$V(x)=V_\mathrm{lattice}(x)+V_\mathrm{ext}(x)$.

In deriving the BHH, one makes very similar assumptions to those made when
discretizing the continuous NLS on a lattice to obtain the DNLS.  Specifically,
both derivations invoke a lowest Bloch band tight-binding approximation.
However, in the latter case, a single configuration of bosons is assumed from
the onset.  The full quantum treatment allows for quantum depletion out of the
condensate mode and thus can describe the system in the strongly interacting
regime.

A general pure state of the full many-body quantum system can be written as a
complex linear superposition of states, each with a well-defined number of
particles in each Wannier state:
\begin{equation}
|\Psi\rangle = \sum_{n_1,n_2,\dots,n_M=0}^{d-1} c_{n_1 n_2\cdots n_M}|n_1 n_2
\cdots n_M\rangle, \label{eqn:genstate}
\end{equation}
where $n_k$ is the number of particles at site $k$.  For obvious computational
reasons, we truncate the local Hilbert space at local dimension $d$, i.e., we
restrict the occupation of each Wannier state to contain at most $d-1$ bosons.
The Hilbert space containing all pure states of the full many-body system is
thus of dimension $d^M$ which becomes prohibitively large for large systems.
For example, even for $d=2$ we can only simulate $M=12$ or 13 lattice sites on
a single PC without further refining the numerical algorithm.  We overcome this
difficulty with use of the time-evolving block decimation routine which will be
discussed in Sec \ref{sec:tebd}.

\subsection{Discrete Mean-Field Theory From Discrete Quantum
Many-Body Theory}

Next, we show how the DNLS can be recovered from the BHH.  The destruction
operator at site $k$ can be evolved in time in the Heisenberg picture according
to $i\hbar\partial_t\hat{b}_k=[\hat{b}_k,\Hhat]$.  After computing the
commutators, we arrive at
\begin{equation}
i\hbar \partial_t \hat{b}_k = -J (\hat{b}_{k+1} + \hat{b}_{k-1}) +
U\hat{b}_k\hat{b}^\dagger_k\hat{b}_k + \epsilon_k\hat{b}_k
\label{eqn:bhatkevolve}.
\end{equation}
We can then take the expectation value of Eq. (\ref{eqn:bhatkevolve}) to obtain
an equation of motion for the order parameter $\langle\hat{b}_k\rangle$.  The
DNLS is recovered exactly if the expectation value is taken with respect to a
product of atom-number Glauber coherent states.  That is, for full many-body
states of the form
\begin{equation}
|\Psi\rangle = \bigotimes_{k=1}^M |z_k\rangle\mathrm{,~where~ }|z_k\rangle=
e^{-\frac{|z_k|^2}{2}}\sum_{n=0}^\infty\frac{z_k^n}{\sqrt{n!}}|n\rangle,
\label{eqn:cohstate}
\end{equation}
we obtain the DNLS for the equation of motion governing the coherent state
amplitude $z_k=\langle\hat{b}_k\rangle$:
\begin{equation}
i\hbar\partial_t z_k = -J\left (z_{k+1} + z_{k-1}\right) + U|z_k|^2 z_k +
\epsilon_k z_k. \label{eqn:dnlsez}
\end{equation}
As discussed in Sec. \ref{sec:qmbt}, for numerical calculations we must
truncate the local Hilbert space to a finite dimension $d$, in which case the
on-site coherent states of Eq. (\ref{eqn:cohstate}) become \emph{truncated}
coherent states:
\begin{equation}
|\Psi\rangle = \bigotimes_{k=1}^M |z_k\rangle\mathrm{,~where~ }|z_k\rangle=
\mathcal{N}_d\, e^{-\frac{|z_k|^2}{2}}
\sum_{n=0}^{d-1}\frac{z_k^n}{\sqrt{n!}}|n\rangle, \label{eqn:truncohstate}
\end{equation}
and $\mathcal{N}_d$ is a normalization factor.

The coherent states of Eq. (\ref{eqn:cohstate}) are known to well-describe the
ground state of the BHH for $J\gg U$ in the limit of an infinite number of
sites $M$ and particles $N$ at fixed filling $N/M$
\cite{Zwerger03_QuantSemiOpt_5_S9}.  It is in this regime that quantum
depletion can be safely neglected and Eq. (\ref{eqn:dnlsez}) is an accurate
description of the system. However, the lattice must still be deep enough so
that the single-band tight-binding approximation is still valid.  In Sec.
\ref{sec:darksolis}, we use the truncated coherent states of Eq.
(\ref{eqn:truncohstate}) to create nonequilibrium initial quantum states in the
BHH that are analogs to the dark soliton solutions of the DNLS.

\section{Numerical Methods}

\subsection{Time-Evolving Block Decimation Algorithm}\label{sec:tebd}

The time-evolving block decimation (TEBD) algorithm was first introduced in
2003-2004 by Vidal \cite{Vidal03_PRL_91_147902,Vidal04_PRL_93_040502} in the
context of quantum computation.  Soon thereafter, Daley \emph{et al.}
\cite{Daley04_JStatMech_P04005} and White and Feiguin
\cite{White04_PRL_93_076401} translated the algorithm into more the familiar
density matrix renormalization group (DMRG) language and showed that TEBD is
equivalent to a time-\emph{adaptive} DMRG routine. Here, we summarize our
implementation of TEBD as applied to the BHH.

\subsubsection{The Vidal Decomposition}

The Vidal prescription is to first rewrite the coefficients in Eq.
(\ref{eqn:genstate}) as a product of $M$ tensors $\{\Gamma^{[\ell]}\}$ and
$M-1$ vectors $\{\lambda^{[\ell]}\}$:
\begin{equation}
c_{n_1 n_2\cdots n_M} = \sum_{\alpha_1,\dots,\alpha_{M-1}=1}^\chi
\Gamma_{\alpha_1}^{[1]n_1}\lamtens{1}\Gamtens{1}{2}\lamtens{2}
\Gamtens{2}{3}\cdots\Gamma_{\alpha_{M-1}}^{[M]n_M}. \label{eqn:vidaldecomp}
\end{equation}
There does exist a procedure for determination of the $\Gamma$s and $\lambda$s
given known coefficients of an arbitrary state; however, this is not generally
useful because one does not typically have access to each component of the
$d^M$-dimensional vector $|\Psi\rangle$.  This procedure would require a
Schmidt decomposition (SD) at every bipartite splitting of the lattice, where
$\chi\leq d^{\lfloor M/2\rfloor}$ is the number of Schmidt vectors retained at
each splitting.  The Schmidt number $\chi_S$, i.e., the number of Schmidt basis
sets required for an exact representation of the state at each cut, is
naturally a measure of global entanglement between the lattice sites
\cite{Vidal03_PRL_91_147902, nielsenMA2000}.  The decomposition
(\ref{eqn:vidaldecomp}) is thus appropriate when $|\Psi\rangle$ is only
slightly entangled according to the Schmidt number, in which case it is
computationally feasible to take $\chi\approx\chi_S$.

\subsubsection{Two-Site Operation}\label{sec:2siteop}

One of the reasons why this decomposition is useful is that it allows for
efficient application of two-site unitary operations. Let us consider a
two-site unitary operation $\hat{V}=\sum V_{n'_\ell n'_{\ell+1}}^{n_\ell
n_{\ell+1}}|n_\ell n_{\ell+1}\rangle\langle n'_\ell n'_{\ell+1}|$ acting on
sites $\ell$ and $\ell+1$.  First, we write $|\Psi\rangle$ in terms of Schmidt
vectors for the subsystems $[1\cdots \ell-1]$ and $[\ell+2\cdots M]$:
\begin{align}
|\Psi\rangle = & \sum_{\alpha_{\ell-1},\alpha_\ell,\alpha_{\ell+1};n_\ell,
n_{\ell+1}}\lamtens{\ell-1}\Gamtens{\ell-1}{\ell}\lamtens{\ell}\Gamtens{\ell}{\ell+1}
|\Phi_{\alpha_{\ell-1}}^{[1\cdots \ell-1]}\rangle\otimes|n_\ell
n_{\ell+1}\rangle\otimes|\Phi_{\alpha_{\ell+1}}^{[\ell+2\cdots M]}\rangle
\nonumber \\ & =
\sum_{\alpha_{\ell-1},\alpha_{\ell+1};n_\ell,n_{\ell+1}}\Theta_{\alpha_{\ell-1}\alpha_{\ell+1}}^{n_\ell
n_{\ell+1}}|\Phi_{\alpha_{\ell-1}}^{[1\cdots \ell-1]}\rangle\otimes|n_\ell
n_{\ell+1}\rangle\otimes|\Phi_{\alpha_{\ell+1}}^{[\ell+2\cdots M]}\rangle
\label{eqn:newPsi}
\end{align}
by invoking Eqs. (13) and (14) of \cite{Vidal03_PRL_91_147902}, where
\begin{equation}
\Theta_{\alpha_{\ell-1}\alpha_{\ell+1}}^{n_\ell
n_{\ell+1}}\equiv\sum_{\alpha_\ell}\lamtens{\ell-1}\Gamtens{\ell-1}{\ell}\lamtens{\ell}
\Gamtens{\ell}{\ell+1}\lamtens{\ell+1} \label{eqn:Theta}
\end{equation}
and $\alpha_\ell\in\{1,2,\dots,\chi\}$.  Note that this definition of the
tensor $\Theta$ differs from an analogous construct in
\cite{Vidal03_PRL_91_147902} which is also denoted $\Theta$ in that work.  We
are up to this point assuming that we know the decomposition
(\ref{eqn:vidaldecomp}) of $|\Psi\rangle$, and hence we also know all elements
of $\Theta$. However, by writing $|\Psi\rangle$ in the form of Eq.
(\ref{eqn:newPsi}) we can easily write the updated state after the application
of $\hat{V}$ as
\begin{equation}
\hat{V}|\Psi\rangle = \sum_{\alpha_{\ell-1},\alpha_{\ell+1};n_\ell,
n_{\ell+1}}\tilde{\Theta}_{\alpha_{\ell-1}\alpha_{\ell+1}}^{n_\ell
n_{\ell+1}}|\Phi_{\alpha_{\ell-1}}^{[1\cdots \ell-1]}\rangle\otimes|n_\ell
n_{\ell+1}\rangle\otimes|\Phi_{\alpha_{\ell+1}}^{[\ell+2\cdots M]}\rangle,
\end{equation}
where $\tilde{\Theta}$ can be written in terms of the updated tensors
$\tilde{\Gamma}^{[\ell]}$ and $\tilde{\Gamma}^{[\ell+1]}$ and the updated
vector $\tilde{\lambda}^{[\ell]}$:
\begin{equation}
\tilde{\Theta}_{\alpha_{\ell-1}\alpha_{\ell+1}}^{n_\ell n_{\ell+1}} =
\sum_{n'_\ell, n'_{\ell+1}}V_{n'_\ell n'_{\ell+1}}^{n_\ell
n_{\ell+1}}\Theta_{\alpha_{\ell-1}\alpha_{\ell+1}}^{n'_\ell n'_{\ell+1}} =
\sum_{\tilde{\alpha}_\ell}\lamtens{\ell-1} \tilde{\Gamma}_{\alpha_{\ell-1}
\tilde{\alpha}_\ell}^{[\ell]n_\ell}\lamtilde{\ell}
\tilde{\Gamma}_{\tilde{\alpha}_{\ell}
\alpha_{\ell+1}}^{[\ell+1]n_{\ell+1}}\lamtens{\ell+1}. \label{eqn:Thetatilde}
\end{equation}
In practice, a given two-site operation is performed as follows: (1) form
$\Theta$ from current $\Gamma$s and $\lambda$s [Eq. (\ref{eqn:Theta})]; (2)
update $\Theta$ by applying $\hat{V}$ to obtain $\tilde{\Theta}$ [Eq.
(\ref{eqn:Thetatilde})]; (3) reshape $\tilde{\Theta}$ from a 4-tensor to a
$(\chi d)\times(\chi d)$ matrix; (4) perform a singular value decomposition
(SVD) on this matrix retaining only the largest $\chi$ singular values
$\tilde{\lambda}_{\alpha_\ell}^{[\ell]}$; and (5) divide out the previous
values of $\lambda^{[\ell-1]}$ and $\lambda^{[\ell+1]}$ in order to compute
$\tilde{\Gamma}^{[\ell]}$ and $\tilde{\Gamma}^{[\ell+1]}$ from the matrix
elements obtained via the SVD. The most expensive computational steps are (1),
the formation of $\Theta$, and (2), the update of $\Theta$ after the
application of $\hat{V}$.  The former requires $\mathcal{O}(d^2 \chi^3)$
elementary operations, whereas the latter requires $\mathcal{O}(d^4 \chi^2)$
elementary operations; hence, our overall two-site operation scales as
$\mathcal{O}[\max( d^2 \chi^3, d^4 \chi^2)]$.

\subsubsection{Real Time Evolution} \label{sec:TEBDtime}

The BHH is a sum of one- and two-site operations, but the terms multiplying $J$
in Eq. (\ref{eqn:bhh}) do not all commute, so the time evolution operator
$e^{-i \Hhat t/\hbar}$ does not directly factor into a product of one- and
two-site unitary operations. However, because the BHH only links nearest
neighbors, we write $\Hhat=\Hhat_\mathrm{odd}+\Hhat_\mathrm{even}$, where
\begin{align}
\Hhat_\mathrm{odd}=&-J \sum_{i~\mathrm{odd}}(\hat{b}_{i+1}^\dagger \hat{b}_{i}
+ \mathrm{h.c.})+\sum_{i~\mathrm{odd}}\left[\frac{U}{2} \hat{n}_i
(\hat{n}_i-\hat{\mathbbm{1}}) + \epsilon_i \hat{n}_i\right]~\mathrm{and}
\\ \Hhat_\mathrm{even}=&-J
\sum_{i~\mathrm{even}}(\hat{b}_{i+1}^\dagger \hat{b}_{i} +
\mathrm{h.c.})+\sum_{i~\mathrm{even}}\left[\frac{U}{2}\hat{n}_i
(\hat{n}_i-\hat{\mathbbm{1}}) + \epsilon_i \hat{n}_i\right].
\end{align}
Each term \emph{within} both $\Hhat_\mathrm{odd}$ and $\Hhat_\mathrm{even}$
commute even though $[\Hhat_\mathrm{odd},\Hhat_\mathrm{even}]\neq 0$.  It is
then convenient to utilize a Suzuki-Trotter approximation of the time evolution
operator for small time steps $\delta t$. Specifically, we employ the
second-order expansion: $e^{-i \Hhat \delta t/\hbar} \approx e^{-i
\Hhat_\mathrm{odd} \delta t/2\hbar} e^{-i \Hhat_\mathrm{even} \delta t/\hbar}
e^{-i \Hhat_\mathrm{odd}\delta t/2\hbar}$, where each exponential factor can be
factored into a product of two-site unitaries. Even though the terms involving
$\hat n$ are one-site operations, we still treat them as two-site operations by
appropriate tensor products with the identity operator. In practice, we build
$d^2$-dimensional matrix representations of $\Hhat$ for each lattice link and
diagonalize these matrices to obtain matrix representations of the two-site
unitary operators. Then, in conjunction with the Suzuki-Trotter expansion, we
employ the two-site operation procedure outlined in Sec. \ref{sec:2siteop} on
an initial decomposed configuration $|\Psi\rangle$ $\mathcal{O}(M)$ times for
each of $t_f/\delta t$ total time steps, updating the decomposition at each
step.  It is straightforward to calculate single-site observables, e.g., the
expectation value of the number operator $\langle\hat{n}_k\rangle$, and
two-site observables, e.g., the one-body density matrix
$\langle\hat{b}^\dagger_i \hat{b}_j\rangle$, by using the partial trace to
calculate the reduced density matrix of the subsystem of interest.  For
example, to calculate $\langle\hat{b}^\dagger_i \hat{b}_j\rangle$, we first
compute $\hat\rho_{ij}=\mathrm{tr}_{k\neq i,j}|\Psi\rangle\langle\Psi|$ using
Eq. (\ref{eqn:vidaldecomp}) for $|\Psi\rangle$ and then use
$\langle\hat{b}^\dagger_i
\hat{b}_j\rangle=\mathrm{tr}(\hat{b}^\dagger\otimes\hat{b}~\hat\rho_{ij})$.
Overall, our implementation of the TEBD algorithm scales as $\mathcal{O}[M
\frac{t_f}{\delta t}\max(d^2 \chi^3, d^4 \chi^2)]$.

\subsubsection{Sources of Error and Convergence Properties}

The TEBD algorithm makes two important approximations: (1) the retention of
only the $\chi$ most heavily weighted basis sets during a given two-site
operation (see Sec. \ref{sec:2siteop}), and (2) the Suzuki-Trotter
representation of the time evolution operator (see Sec. \ref{sec:TEBDtime}).
For the latter case, we find that for the results presented in Sec.
\ref{sec:darksolis} it is sufficient to use time steps of size $\delta t =
0.01\,\hbar/J$ to obtain converged results.  The former approximation is more
subtle as its accuracy is directly related to the amount of entanglement
present in the system.  Specifically, in Sec. \ref{sec:darksolis}, we time
evolve mean-field initial states [see Eq. (\ref{eqn:truncohstate})] for which
$\chi=1$ is sufficient for exact representation; however, unitary time
evolution increases entanglement between sites.  To ensure that our choice of
$\chi$ is sufficient, we run equivalent simulations with increasing values of
$\chi$ and look for convergence of calculated observables, e.g., average local
number $\langle\hat{n}_k\rangle$. It is important to point out that, owing to
the local nature of the expansion (\ref{eqn:vidaldecomp}), accurate calculation
of nonlocal observables, e.g., off-diagonal elements of the single-particle
density matrix, converge more slowly with respect to $\chi$.  For the
observables and time scales presented in Sec. \ref{sec:darksolis}, our results
are converged for the specified values of $\chi=45,50$.  We also note that the
fidelity of truncation at $\chi$ eigenvalues can be quantified by the sum of
the non-retained eigenvalues after application of a two-site operation. This
quantity can be interpreted as a measure of the amount of entanglement
generated by the two-site operation. Typically, for a single two-site unitary,
we find this truncation error to be less than or on the order of $10^{-6}$.

\subsection{Constrained Imaginary Time Relaxation in DNLS}

\subsubsection{Fundamental Dark Soliton Solutions} \label{sec:standing}

In Sec. \ref{sec:darksolis}, we use the TEBD routine to simulate the quantum
evolution of the dark soliton solutions of the DNLS by using truncated coherent
states as initial configurations. This requires knowledge of the set of
coherent state amplitudes $\{z_k\}$ corresponding to a discrete dark soliton.
Using a standard Crank-Nicolson scheme for the time-stepping procedure, we
calculate the standing dark soliton solution of the DNLS by performing
constrained imaginary time relaxation on Eq. (\ref{eqn:dnlsez}) with
$\epsilon_k=0$.  Specifically, we take an initial condition of form $z_k =
mx_k$ where $x_k$ is the position of the $k$th site and $x=0$ is the center of
the lattice and normalize the solution to $N_\mathrm{DNLS}=\sum_{k=1}^M|z_k|^2$
at each step of imaginary time. The stationarity of the solution is tested by
subsequent evolution in real time.

\subsubsection{Density and Phase Engineering of Gray Solitons} \label{sec:solieng}

We also consider the case of two solitons moving toward one another at finite
velocity.  These initial conditions are obtained via the methods of density and
phase engineering for soliton creation \cite{Carr01_PRA_63_051601} as applied
to the DNLS.  We first perform imaginary time relaxation on a uniform initial
condition with an external potential of the form
\begin{equation}
\epsilon_k = V_0\left\{\exp{\left[-\frac{(x_k+\xi)^2}{2
\sigma_\epsilon^2}\right]}+ \exp{\left[-\frac{(x_k-\xi)^2}{2
\sigma_\epsilon^2}\right]}\right\} \label{eqn:gausspot}
\end{equation}
to dig two density notches each centered at distance $\xi$ from the center of
the lattice.  Next, we imprint an instantaneous phase of the form
\begin{equation}
\theta_k = \Delta\theta\left
\{-\frac{1}{2}\tanh\left[\frac{2(x_k+\xi)}{\sigma_\theta}\right]
+\frac{1}{2}\tanh\left[\frac{2(x_k-\xi)}{\sigma_\theta}\right]+1 \right\}
\label{eqn:phase}
\end{equation}
which gives the solitons equal-and-opposite initial velocities toward the
center of the lattice.  Phonon generation is minimized by appropriately tuning
the width $\sigma_\theta$ of the phase profiles to the soliton depth as
determined by $V_0$ in the density engineering stage.

\section{Time Evolution of Quantum Solitons}\label{sec:darksolis}

With an initial DNLS configuration $\{z_k(0)\}$ obtained either via the
procedure outlined in Sec. \ref{sec:standing} for a single standing soliton or
the procedure outlined in Sec. \ref{sec:solieng} for two colliding solitons, we
then build a product of truncated coherent states according to Eq.
(\ref{eqn:truncohstate}) for input into the TEBD quantum simulation routine.
The Vidal decomposition (\ref{eqn:vidaldecomp}) of a product state
$|\Psi\rangle=\bigotimes_{k=1}^M\left(\sum_{n_k=0}^{d-1}c_{n_k}^{(k)}|n_k\rangle\right)$
is trivial to compute:
\begin{equation}
\lamtens{\ell}=\kron{\alpha_{\ell}}{1}~~\mathrm{and}~~\Gamtens{\ell-1}{\ell}
=c_{n_\ell}^{(\ell)} \kron{\alpha_{\ell-1}}{1} \kron{\alpha_\ell}{1},
\end{equation}
where for the case of truncated coherent states $c_n^{(k)}=\mathcal{N}_d \,
e^{-|z_k|^2/2}\frac{z_k^n}{\sqrt{n!}}$.

An extensive discussion of results obtained using the above methodology to
create nonequilibrium dark soliton initial states in the BHH is presented in
Refs. \cite{Mishmash07_short, Mishmash08_long}.  We will summarize those
results here.  Being direct analogs of mean-field solitons, the initial
conditions analyzed below do not conserve total particle number, although
quantum evolution does conserve total \emph{average} particle number. That is,
in this section, we consider the quantum many-body evolution of mean-field-like
solitons and refer to these structures as \emph{quantum solitons}.  However, we
stress that neither the discrete mean-field theory (DNLS) nor the corresponding
quantum theory (BHH) are integrable systems, so these are not solitons in the
mathematically rigorous sense. It is possible to density and phase engineer
dark soliton states directly in the BHH that are eigenfunctions of the total
number operator. In Ref. \cite{Mishmash08_long}, we use a number-conserving
version of the TEBD routine to generate and analyze the quantum dynamics of
such states. Although some observables behave differently in this case, the
conclusions reached are generally the same.

\subsection{Standing Solitons}

\newcommand{\plotscale}[0]{0.31}

\begin{figure}[t]
\begin{center}
\subfigure{\scalebox{\plotscale}{\includegraphics[angle=0]{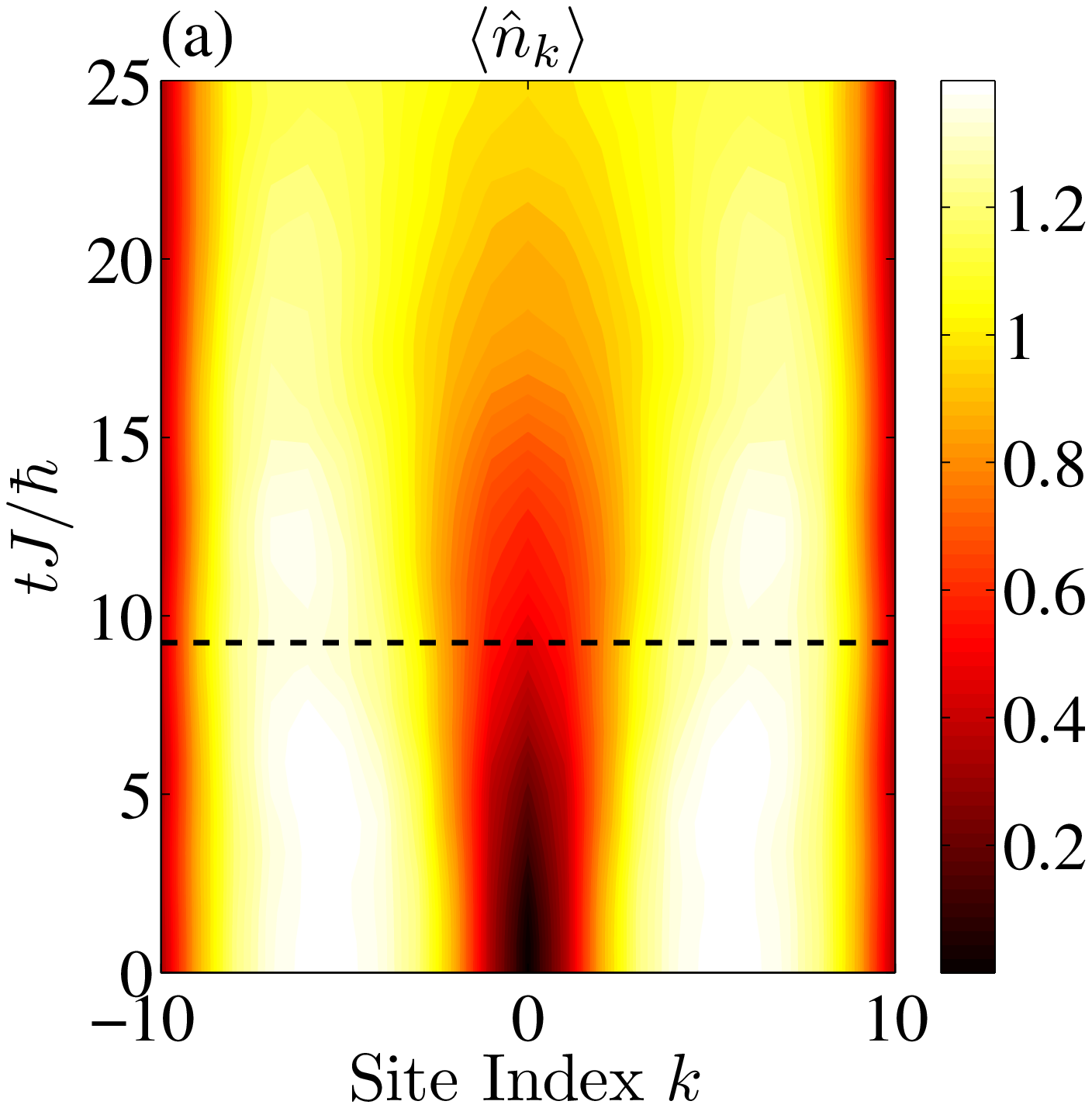}}}
\subfigure{\scalebox{\plotscale}{\includegraphics[angle=0]{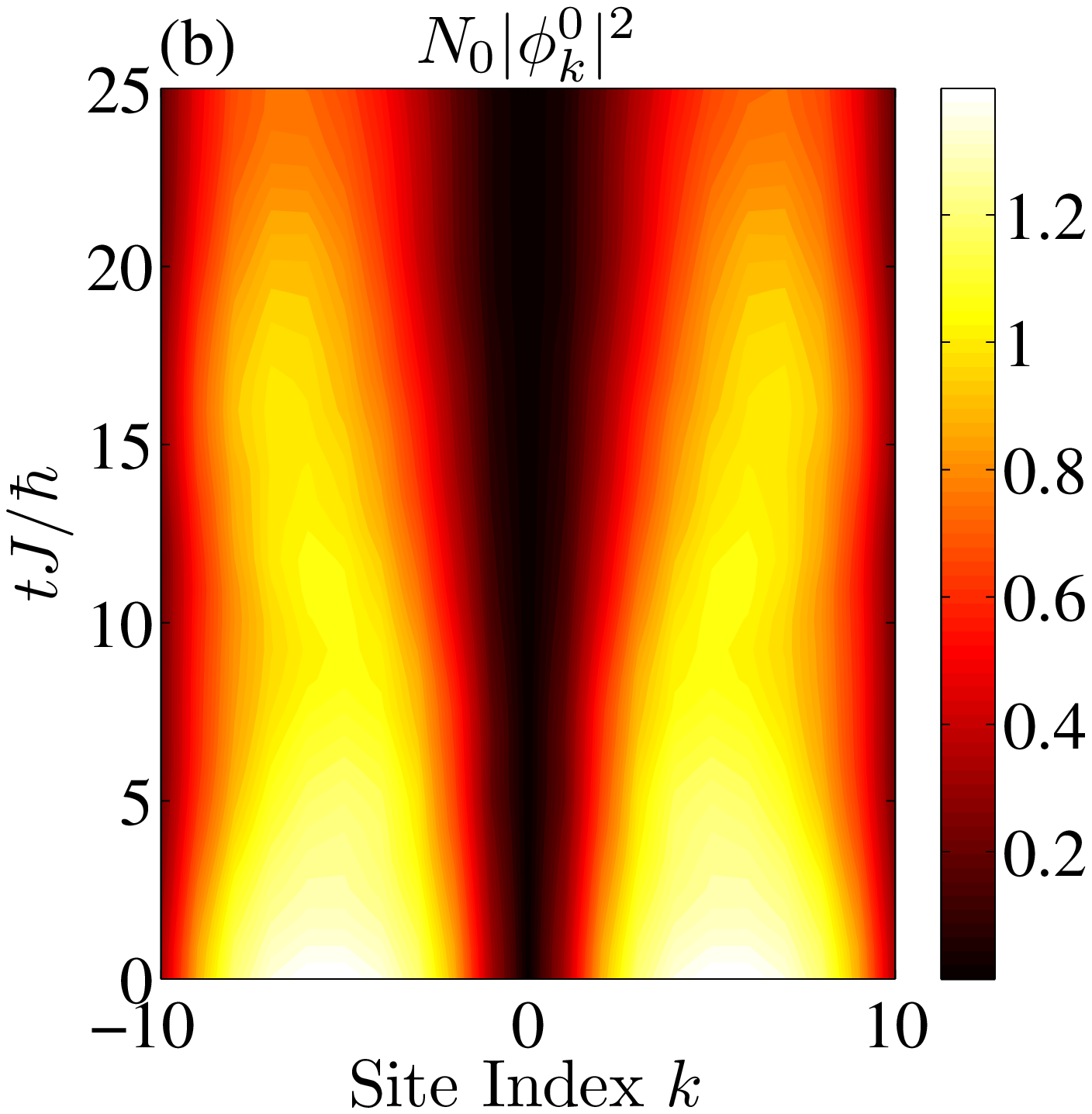}}}
\subfigure{\scalebox{\plotscale}{\includegraphics[angle=0]{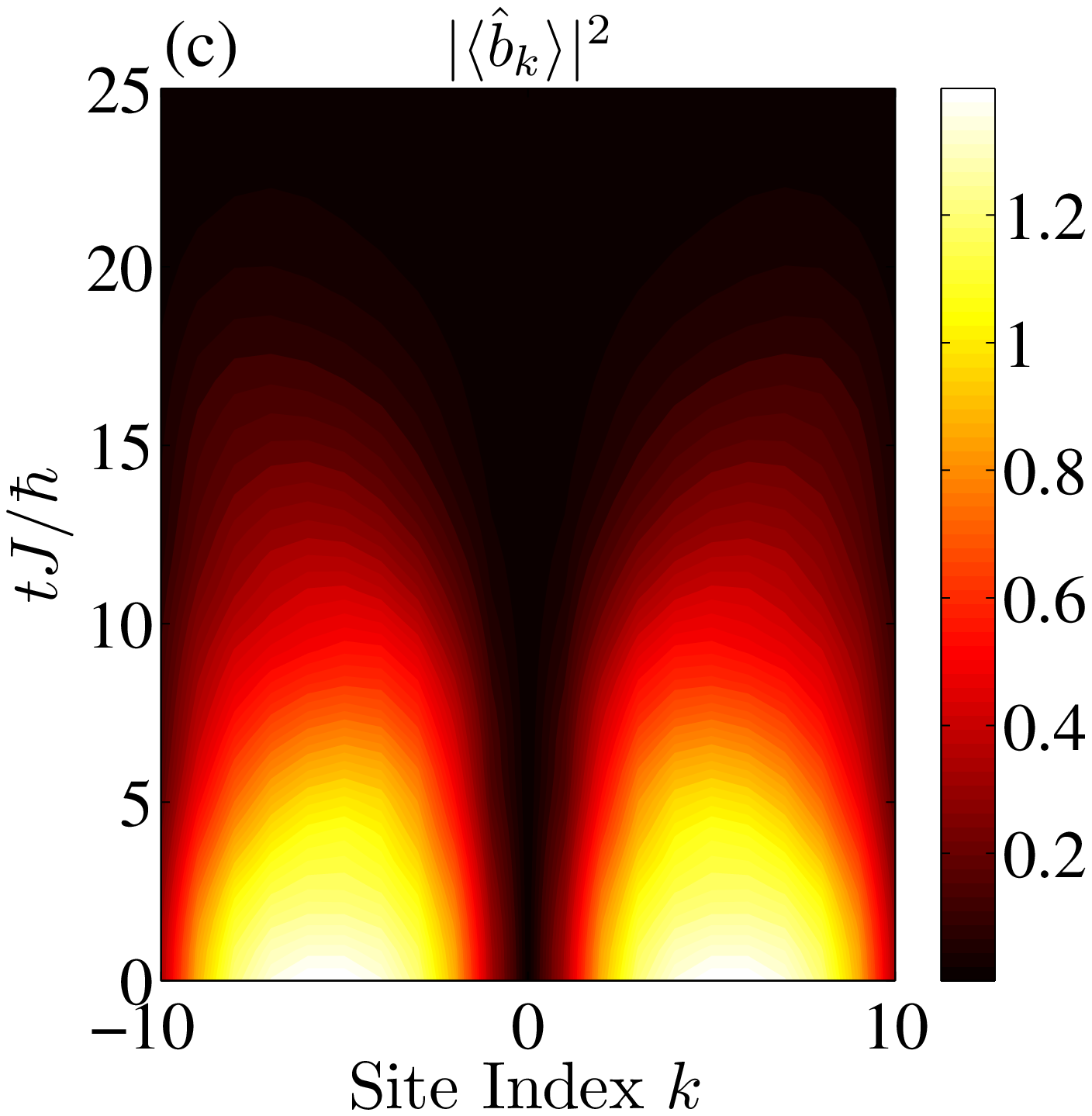}}}
\end{center}
\caption{\emph{Density measures for a standing quantum soliton.} Quantum
evolution of (a) average particle number, (b) condensate wave function
\cite{Penrose56_PR_104_576}, and (c) order parameter versus position and time
for a standing DNLS dark soliton initial configuration. The dashed line in (a)
indicates the $1/e$ decay time of the order parameter norm $N_b$.
 \label{fig:standing}}
\end{figure}

The DNLS assumes a single configuration of bosons, i.e., bosons are only
allowed to occupy one single-particle orbital.  However, for an $M$ mode
system, a full quantum treatment will permit bosons to occupy any of the $M$
permissible modes.  For the system sizes accessible to explore with the TEBD
routine, a standing soliton initial DNLS configuration exhibits a finite
lifetime due to quantum effects indescribable by mean-field theory.  Most
notably, quantum evolution causes significant quantum depletion out of the
initial dark soliton configuration into higher order orbitals which fill in the
soliton density notch, where as discussed in \cite{Mishmash07_short}, the
natural orbitals of the system are defined as the eigenfunctions of the
one-body density matrix $\langle\hat{b}_i^\dagger\hat{b}_j\rangle$
\cite{Penrose56_PR_104_576}. We find that the soliton lifetime is closely
correlated to the the growth in quantum effects such as a decay in the order
parameter norm $N_b=\sum_{k=1}^M |\langle\hat b_k\rangle|^2$ and growth in the
generalized entropy $Q = \frac{d}{d-1} \left[1 - \frac{1}{M}\sum_{k=1}^M
\mathrm{tr}(\hat{\rho}_k^2) \right]$. Shown in Fig. \ref{fig:standing} is the
evolution of density measures of a quantum soliton with parameters $\nu
U/J=0.35$, $\nu=1$, $M=21$, $d=7$, $\chi=45$, where $\nu=N_\mathrm{DNLS}/M$ is
approximately equal to the average filling.

\subsection{Soliton-Soliton Collisions}

\begin{figure}[H]
\begin{center}
\subfigure{\scalebox{\plotscale}{\includegraphics[angle=0]{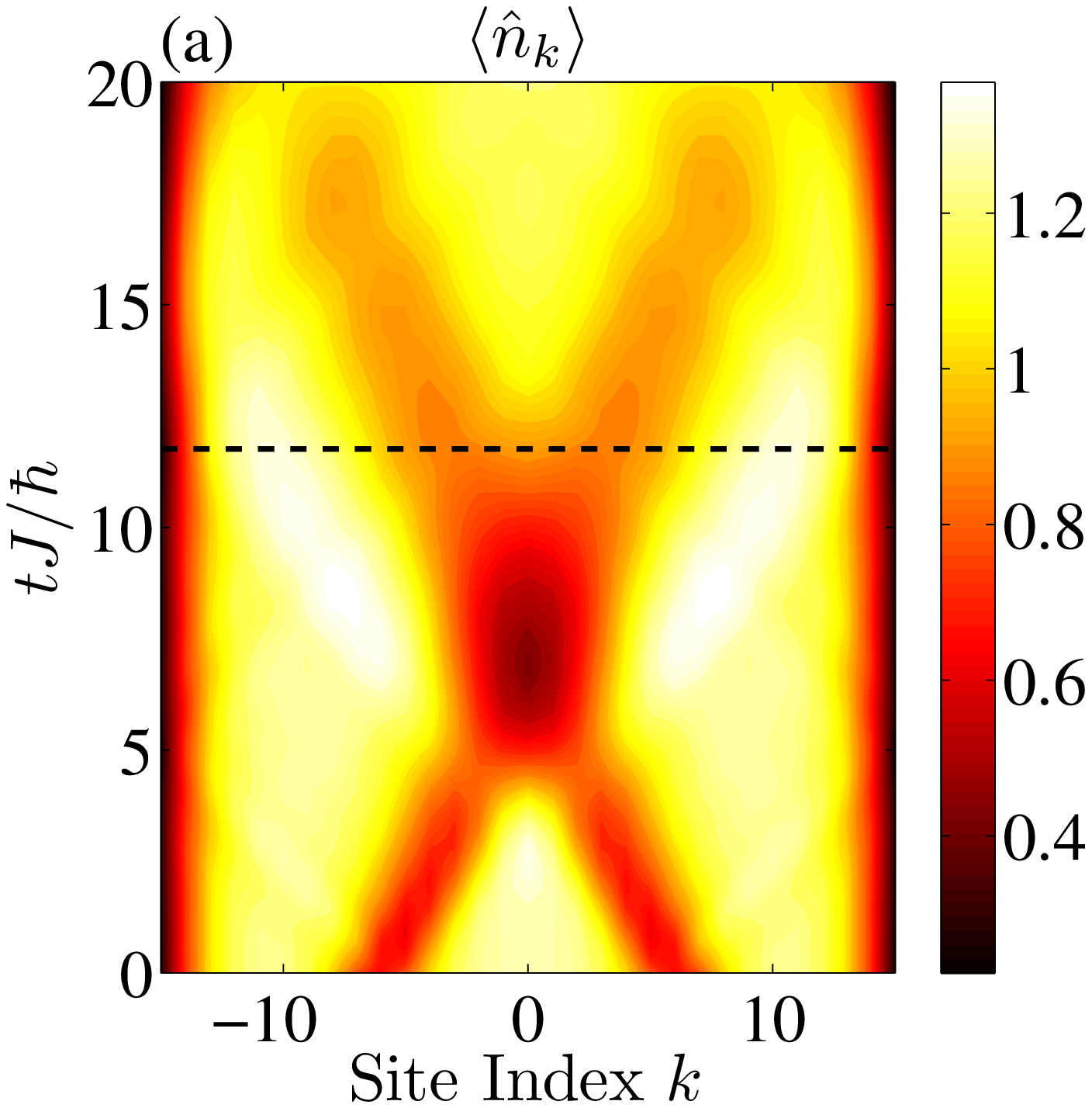}}}
\subfigure{\scalebox{\plotscale}{\includegraphics[angle=0]{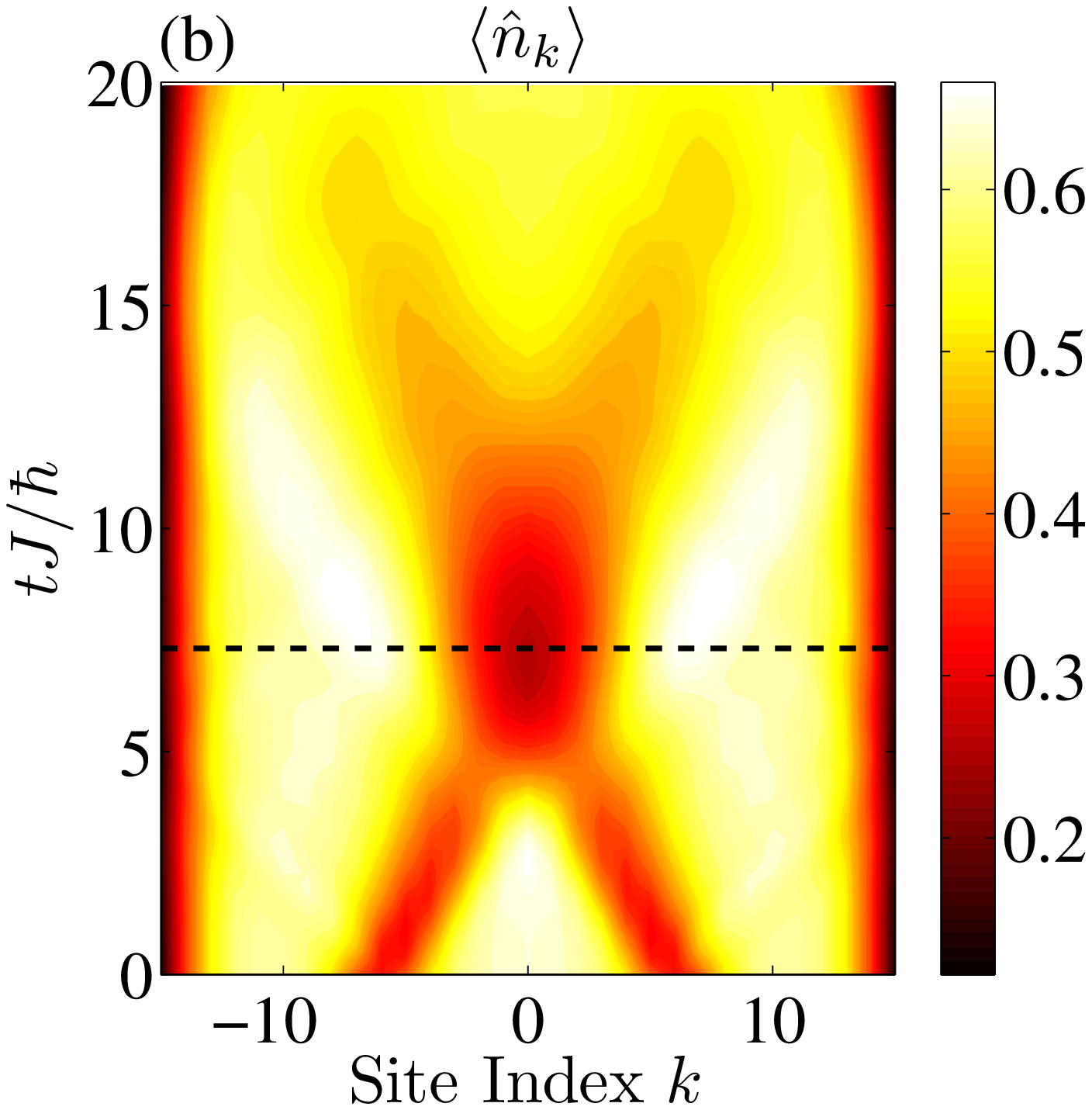}}}
\subfigure{\scalebox{\plotscale}{\includegraphics[angle=0]{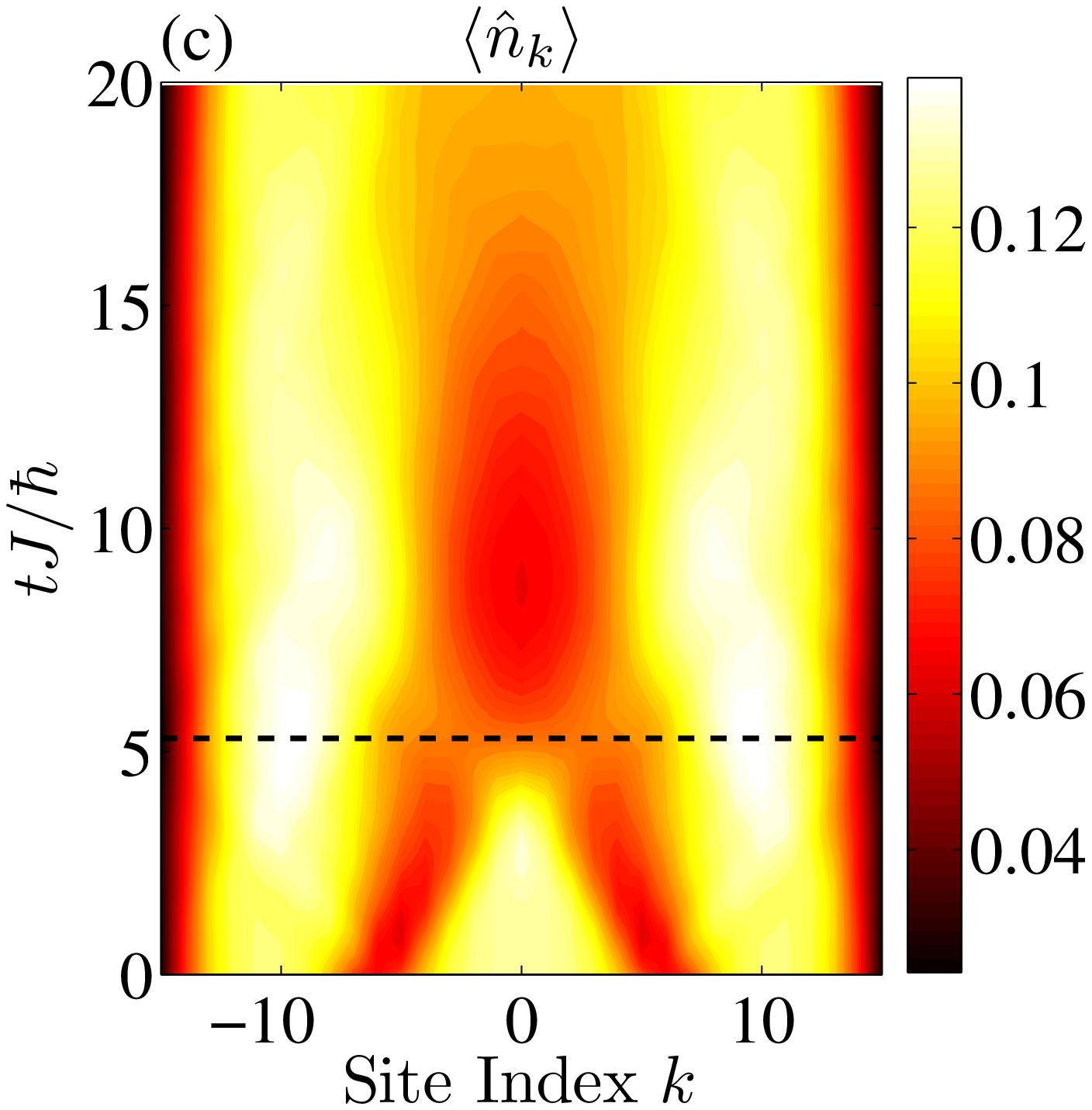}}}
\end{center}
\caption{\emph{Quantum soliton collisions and decoherence-induced
inelasticity.}  Average particle number for two colliding quantum solitons at
filling factors (a) $\nu=1$, (b) $\nu=0.5$, and (c) $\nu=0.1$.  The collision
elasticity is decreased when the decoherence time (dashed lines) occurs at or
before the time of collision.  \label{fig:colliding}}
\end{figure}

In Fig. \ref{fig:colliding}, we display the quantum evolution of two colliding
dark solitons where the initial conditions were obtained by density and phase
engineering in the DNLS as summarized in Sec. \ref{sec:solieng}.  Here, if the
decoherence time, as measured by the decay time of the order parameter norm,
occurs before or near the collision time, then there is a loss in elasticity of
the soliton collision.  For a fixed value of the effective nonlinearity $\nu
U/J$, we can independently tune the decoherence time by changing the filling
$\nu$ without altering the initial density-phase profile of the solitons. In
Fig. \ref{fig:colliding}, we depict this effect in three separate simulations
with parameters $\nu U/J=0.35$, $M=31$, $\chi=50$, $V_0/J=0.4$,
$\sigma_\epsilon/a=1$, $\Delta\theta=0.3\pi$, $\sigma_\theta/a=2$, $\xi/a=6$ at
filling factors $\nu=1,0.5,0.1$, where $a$ is the lattice constant.

\section{Conclusion}

In conclusion, we have summarized the derivations of the DNLS both from the
continuous mean field and from the continuous quantum field.  In both cases, we
invoke a single-band tight-binding approximation; however, the latter
derivation from the quantum many-body perspective is more insightful as it is
based on single-particle physics with single-orbital occupation not assumed
until the end.  Using Vidal's TEBD routine to propagate dark soliton DNLS
configurations forward in time according to the BHH, we have shown that quantum
fluctuations give dark solitons a finite lifetime and induce an inelasticity in
soliton-soliton collisions.  For a more extensive analysis of these results, we
refer the reader to Refs. \cite{Mishmash07_short, Mishmash08_long}.

We thank Charles Clark, Ippei Danshita, and Jamie Williams for useful
discussions.  This material is based upon work supported by the National
Science Foundation under Grant No. PHY-0547845, as part of the NSF CAREER
program.

% \appendix

% \section{}
% \label{}

%For final submission, we will have to type out the bibliography,
%but for now we can just use the Phys. Rev. style...
%\bibliographystyle{prsty}
%\bibliography{../../References/00Database}

\end{document}